\documentclass[runningheads]{llncs}
\usepackage[T1]{fontenc}
\usepackage{graphicx}
\usepackage{hyperref}
\usepackage{color}

\usepackage{subcaption}
\usepackage{amssymb,amsmath}
\usepackage{adjustbox}
\usepackage{tikz}
\renewcommand{\theta}{\vartheta}
\renewcommand{\phi}{\varphi}
\renewcommand{\epsilon}{\varepsilon}

\usepackage{placeins} %
\usepackage{float} %
\usepackage{enumitem}%
\usepackage{bm}
\usepackage{multirow}
\newcommand{\bftab}{\fontseries{b}\selectfont} %
\usepackage[subtle]{savetrees}
\usepackage{mwe} %

\usepackage{xcolor}

\renewcommand{\theta}{\vartheta}

\newcommand{\figureref}[1]{\hyperref[#1]{Figure~\ref{#1}}}
\newcommand{\tableref}[1]{\hyperref[#1]{Table~\ref{#1}}}
\newcommand{\sectionref}[1]{\hyperref[#1]{Section~\ref{#1}}}
\newcommand{\appendixref}[1]{\hyperref[#1]{Appendix~\ref{#1}}}

\renewcommand{\eqref}[1]{\hyperref[#1]{Equation~(\ref{#1})}}

\begin{document}
\title{Modeling the Neonatal Brain Development Using Implicit Neural Representations}
\titlerunning{Modeling the Neonatal Brain Development Using INRs}
\author{
Florentin Bieder\inst{1}\orcidID{0000-0001-9558-0623}    \email{florentin.bieder@unibas.ch}\and
Paul Friedrich\inst{1}\orcidID{0000-0003-3653-5624}      \email{paul.friedrich@unibas.ch}\and
Hélène Corbaz\inst{1}\orcidID{0009-0006-9716-5136}       \email{helene.corbaz@unibas.ch}\and
Alicia Durrer\inst{1}\orcidID{0009-0007-8970-909X}       \email{alicia.durrer@unibas.ch}\and
Julia Wolleb\inst{1}\orcidID{0000-0003-4087-5920}        \email{julia.wolleb@unibas.ch}\and
Philippe C. Cattin\inst{1}\orcidID{0000-0001-8785-2713}  \email{philippe.cattin@unibas.ch}
}

\authorrunning{F. Bieder et al.} %
\institute{ Department of Biomedical Engineering, University of Basel %
}
\maketitle              %

\begin{abstract}
The human brain undergoes rapid development during the third trimester of pregnancy. 
In this work, we model the neonatal development of the infant brain
in this age range.
As a basis, we use MR images of preterm- and term-birth neonates 
from the developing human connectome project (dHCP).
We propose a neural network, specifically an implicit neural representation (INR), to predict
2D- and 3D images of varying time points.
In order to model a subject-specific development process, it is necessary to disentangle
the age from the subjects' identity in the latent space of the INR.
We propose two methods, Subject Specific Latent Vectors (SSL) and Stochastic Global Latent Augmentation (SGLA), enabling this disentanglement.
We perform an analysis of the results and compare our proposed model 
to an age-conditioned denoising diffusion model
as a baseline. We also show that our method can be applied in a memory-efficient way,
which is especially important for 3D data.
The code is available on \href{https://github.com/FlorentinBieder/Neonatal-Development-INR}{https://github.com/FlorentinBieder/Neonatal-Development-INR}.
\end{abstract}


\section{Introduction}

The development of the central nervous system begins 
early in pregnancy and can be detected
using ultrasound as early as eight weeks of 
gestation \cite{monteagudo2007ultrasound}.
At around 20~weeks, the corpus callosum is fully developed, 
and the first structures of the cortical surface begin to form 
\cite{rubenstein2013patterning}.
The process of cortical folding, or gyrification, continues well after term birth.
Across most individuals, the gyrification is similar on a macroscopic level for the major structures,
but  differs on the level of the smaller sulci, even for monozygotic twins  \cite{bozek2018construction}.
Before 20~weeks postmenstrual age (PMA) their brains' appearance  have a high degree of similarity, but then progressively develop an individual character \cite{white2010development}.
In this work, we try to model the subject specific development of the preterm and term neonatal
brain, based on the dHCP dataset (see \sectionref{sec:dhcp}), 
containing brain MR images of neonates 
between 26 and 45 weeks PMA.
This dataset poses challenges due to limited data availability, with each subject 
having scans from at most two or three different points in time, and the majority 
having only a single scan available.

\begin{figure}[htbp]
\begin{center}
\includegraphics[width=0.8\linewidth, clip, trim=0mm 0mm 0mm 0mm]{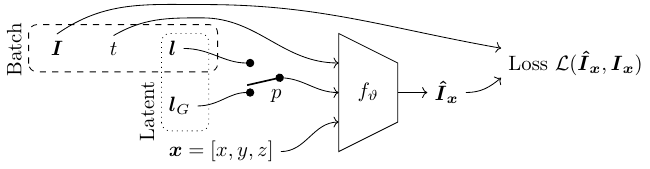}
\end{center}
\caption{Overview: 
  The input to the network $f_\theta$  consists of 
  the spatial coordinates $\vec x$ of the desired output pixel $\vec{\hat I}_{\vec x}$,
  the desired PMA $t \in T$, and 
  a latent vector $\vec l \in \Lambda$, encoding the subject identity. 
  The switch with probability $p$ between
  $\vec l$ and the global latent vector $\vec l_G$ represents the SGLA.}
\label{fig:overview}
\end{figure}

Using an implicit neural representation (INR), we want to  predict the healthy brain
development from a given scan at a given PMA. That is, we want to predict
an image of the same subject at a later or earlier point in time.
Therefore, we have to find an age-agnostic representation of the identity of the subject,
meaning we need to disentangle the age from the identity to be able to represent
the same subject at different points in time. Modelling the development of the healthy brain could then serve as a basis for further downstream tasks, e.g.\ detection of abnormal brain development.
INRs work by representing images (and other signals) as networks
that take the coordinates of a pixel as input and output the corresponding intensities.
This is in contrast to convolutional neural networks (CNNs) or transformers,
which process an entire image, i.e., a grid of pixels, at once.
INRs have been successfully applied to the representation 
and modelling shapes \cite{chen2019learning,mescheder2019occupancy},
images \cite{sitzmann2020implicit}, and recently also natural 3D scene reconstruction
\cite{shen2022nerp}.
In the medical domain, they have been used for various tasks, such as image segmentation \cite{stolt2023nisf}, 
shape completion \cite{amiranashvili2022learning}, 
tomographic reconstruction \cite{shen2022nerp,fang2022snaf},
and image registration \cite{sun2022mirnf}.
\paragraph{Contribution}
We show that an INR can be trained to model the neonatal brain
development based on sparsely- and highly irregularly sampled data with respect to 
the time axis. \figureref{fig:overview} provides an overview
of the model.
To enable the disentanglement along the time axis, we propose the
following two methods that can be applied independently during training:
\begin{itemize}[noitemsep, topsep=0cm]
\item A method to disentangle the subject's PMA at the time 
of the scan from its identity by enforcing a 
subject-specific latent space (SSL).
\item An augmentation method with a global latent vector in the latent space. We call it stochastic global latent augmentation (SGLA).
This performs similarly to SSL but is intended to 
make better use of subjects
with only a single scan in the dataset.
\end{itemize}
We show how SSL and SGLA can improve the disentanglement, and with that, the predictions.
Furthermore, we show that our INR approach can be run on hardware with limited GPU memory.

\section{Method}\label{sec:method}
This work presents an approach based on an INR.
This means that we model our images 
$\vec I \in \mathcal I = \{I:\Omega \to [0,1]\}$ 
at different points of time $t \in T$ as a function
$f_\theta : \Omega \times T \times \Lambda \to [0, 1]$
parametrized by $\theta$, 
where $\Omega \subset \mathbb R^d$ is the spatial image domain,
$T = [0, 1]$ is the time domain (i.e.\ normalized PMA), and
$\Lambda \subset \mathbb R^\lambda$ is our latent space.
For any image $\vec I \in \mathcal I$ we denote the intensity
at $\vec x \in \Omega$ as $\vec I_{\vec x}$.
In the forward pass, for each pixel, 
the normalized input coordinates $\vec x \in \Omega$ are concatenated
with the latent vector $\vec l$ 
and the corresponding PMA $t$ and then passed to the network.
Therefore, during training, for some given
image $\vec I$ with corresponding PMA $t$ and latent vector
$\vec l$, and some input coordinates $\vec x$,
we predict $\hat{\vec I}_{\vec x} := f_\theta(\vec x, t, \vec l)$, and we optimize the $\ell^2$-loss 
\scalebox{0.9}{\smash{
${\mathcal L(\hat{\vec{I}}_{\vec x}, \vec{I}_{\vec x}) = \Vert \hat{\vec I}_{\vec x} - \vec{I}_{\vec x}\Vert_2^2}$
}}
with respect to $\theta$ and $\vec l$. 
This is described in more detail in \sectionref{sec:training}.
For inference, we can project an input image into the latent space by
optimizing the latent vector with respect to the reconstructed image.
To then predict images \scalebox{0.9}{\smash{$\hat{\vec I}^i$}} for different points of time $t_i$, 
for a given latent vector $\vec l$, we can sample 
\scalebox{0.9}{\smash{$\hat{\vec I}_{\vec x}^i := f_\theta(\vec x, t_i, \vec l)$}}
over the whole 
image domain $\vec x \in \Omega$ for every $t_i$. 
This is described in detail in \sectionref{sec:inference}.
The architecture of the network $f_\theta$ is described in \appendixref{appendix:detailedarchitecture}.
\subsection{Age Disentanglement in the Latent Space}
In our setup, the latent space represents the identity of a subject. 
We want to disentangle the identity of the subject from its age.
To this end, we propose the following two methods, which can be employed independently of each other:

\paragraph{Subject Specific Latent Vectors (SSL)}
Previously developed INRs for similar tasks use one latent vector per input image 
\cite{amiranashvili2022learning,stolt2023nisf,chen2019learning,mescheder2019occupancy}.
This is sufficient if we want our network to model a specific input image. 
In our case, we want the model to generalize the
representation of a specific subject across the time domain $T$
and to encode the subject's identity independently, i.e., disentangle
the identity from the age.
Given the availability of images from multiple time points 
for one subject, we propose using the \emph{same} latent vector
for all images of the \emph{same} subject, with 
the goal of forcing the model to use only the age inputs $t \in T$
to encode the age of the subject and use the latent vector
for encoding the identity of the subject only.
\paragraph{Stochastic Global Latent Augmentation (SGLA)}
The second method we propose for disentangling the subjects' age from their identity
is based on a specific type of data augmentation
during training:
We introduce an additional \emph{global} latent vector $\vec l_G$ for training.
In each iteration of the stochastic gradient descent,
the latent vector of the current batch is replaced with this
global latent vector $\vec l_G$ with some probability $p$.
It is intended to mimic SSL for those subjects
that have only a \emph{single} scan in the dataset: 
Similarly as in SSL, the network is trained to predict scans
at different developmental stages for the same latent vector.
Instead of a subject specific latent vector, we use the global vector $\vec l_G$ instead.
This forces the network to take the age input $t$ into account.
We set the probability $p = 10\%$ to make it
comparable to SSL, as about $10\%$ of the subjects in the
training set have more than one scan. We have performed an ablation with respect
to $p$ in \appendixref{appendix:ablationp}.

\subsection{Training \& Inference}\label{sec:training}
The network, along with the latent vectors of the training
set, is jointly optimized using the \emph{AdamW}
optimizer \cite{loshchilov2017decoupled} with a learning rate of 
$\text{lr} = 10^{-4}$ and 
an $\ell _2$-loss.
\label{sec:pixelsampling}
Since the INRs can be processed pixel by pixel, we
do not necessarily need to use the whole image for each
training step.
In our model, we can adjust the percentage of
pixels that we randomly sample from a given image in each training step 
by a hyperparameter. 
We chose to use $5\%$ for our experiments to fit the training of the 
3D INRs on a GPU with 12GB memory. For simplicity,
we used identical settings for the 2D case. %
This lets us easily adjust for the available memory size of a given GPU.
The details of the pixel sampling are described in \appendixref{appendix:pixelsampling}

If we want to use the entire image (be it for training or inference),
we can still profit from the pixel-wise evaluations:
We can serialize the whole input into micro-batches that
are processed sequentially and accumulate the gradients
with a constant amount of memory. We use this for the inference
of the 3D images. The amount of memory used is then mainly
dictated by the micro-batch size.
These methods allow us to train models on devices
with very little memory. We discuss the resource
consumption in \sectionref{sec:resources}.

\paragraph{Inference}\label{sec:inference}
During inference, we use the same loss as during training,
but we keep the network fixed and only optimize the 
latent vector $\vec l$. 
Therefore, for a given image $\vec I^1$ with age $t_1$ as input,
we optimize a latent vector $\vec l$ to minimize the difference between
reconstruction \scalebox{0.9}{\smash{$\hat{\vec I}_{\vec x}^1 = f_\theta(\vec x, t_1, \vec l)$}}
and the image \scalebox{0.9}{\smash{$\vec I_{\vec x}^1$}}
over all $\vec x \in \Omega$, i.e. 
\scalebox{0.9}{\smash{
${\hat{\vec l} := \arg\min_{\vec l \in \Lambda} \sum_{\vec x \in \Omega} \Vert f_\theta(\vec x, t_1,  \vec l) - \vec I_{\vec x}^1\Vert_2^2}$.
}}
We then use this vector $\hat{\vec l} \in \Lambda$ 
to generate a prediction
$\hat{\vec I}$ for some desired age $t_2 \in T$ by computing 
$\hat{\vec I}_{\vec x} = f_\theta(\vec x, t_2, \hat{\vec l})$.
The latent vectors are initialized with zeros and then optimized over $1000$ step for the 3D case and $2000$ for the 2D case,
using \emph{AdamW} with a learning rate of $\operatorname{lr} = 10^{-3}$.

\subsection{Baseline: Denoising Diffusion Models}\label{sec:ddim}
As a baseline, we use a denoising diffusion model with
gradient guidance (DDM+GG) \cite{wolleb2022swiss} to predict
an image for a certain PMA, given some image of a different PMA.
This method has been implemented for 2D images and has been shown to perform well on similar tasks
such transforming portraits into younger or older versions,
and simulating tumor growth over time.

DDM + GG uses two networks in tandem: the denoising and the regression network.
The age conditioning is performed using the gradients of a regression network
during the denoising process, which is trained to predict the subject's PMA.
The denoising and the regression networks were trained 
to convergence (1M and 150k iterations, respectively) with $T=1000$ noising and denoising steps.
The number of noising and denoising steps for inference was $L=600$ 
and the gradient scale $c = 3\cdot 10^5$. 
We optimized both parameters $L$ and $c$ on the test set to make
a fair comparison. Since this method was proposed for 2D images,
we also trained our INR on the same 2D data (axial slices).
The details are reported in \appendixref{appendix:ddim}.

\section{Experiments \& Results}\label{sec:experiments}
In the following section, we report the results of our experiments.
If not otherwise noted, the setup is as follows:
From each subject in the test set, we consider two scans that
were made at a different point of time, i.e., 
at a different PMA of the subject.
As explained in \sectionref{sec:inference}, for a given subject, we first
determine the latent vector $\hat{\vec l}$ based on the input 
image \scalebox{0.9}{\smash{$\vec I^1$}} and PMA $t_1$, that is, we
optimize $
\scalebox{0.9}{\smash{
${\hat{\vec l} := \arg\min_{\vec l \in \Lambda} \sum_{\vec x \in \Omega} \Vert f_\theta(\vec x, t_1,  \vec l) - \vec I_{\vec x}^1\Vert_2^2}$.
}}$
We then use this latent vector $\hat{\vec l}$ to generate a prediction \scalebox{0.9}{\smash{$\hat{\vec I}^2$}}
for the PMA $t_2$ of the second scan $\vec I^2$
by computing \scalebox{0.9}{\smash{$\hat{\vec I}_{\vec x}^2 := f_\theta(\vec x, t_2, \hat{\vec l})$}} for every $x \in \Omega$.
We then compare the predicted image \scalebox{0.9}{\smash{$\hat{\vec I}$}} with the 
second scan \scalebox{0.9}{\smash{$\vec I^2$}}, which serves as the ground truth 
for all metrics that we will introduce below.
This allows us to quantify how well our model predicts the development 
process of the brain.
To justify our contributions, we perform an ablation of our proposed
method, that is, we conduct our experiments with and without
SGLA and SSL, respectively, in 2D as well as in 3D.

\paragraph{Dataset}\label{sec:dhcp}
We used the dHCP (third data release) 
dataset for our experiments  \cite{makropoulos2018developing,bozek2018construction}.
It contains $T1$- and $T2$-weighted neonatal MR-scans
of 329 subjects.
We use the $T2$-weighted scans as these are preferred for
assessing brain structure in fetal and neonatal MRI \cite{makropoulos2018developing},
due to the immature myelination. %
The postmenstrual age (PMA, in weeks) is available for each scan,
ranging from $26$ to $45$ weeks with a median of $40.57$.
Our preprocessing is described in \appendixref{appendix:preprocessing}.

\paragraph{Ablation of SSL and SGLA and comparison to the Baseline}
To compare our predictions on the test set with the corresponding ground truths,
we compute the peak-signal-noise-ratio (PSNR), 
the structural similarity index (SSIM)
and the mean absolute error (MAE).
We report the same metrics for the baseline DDM+GG in \tableref{table:scores}.
Our proposed 2D INR with SGLA and SSL outperforms DDM+GG with respect to every metric.
Furthermore, the performance decreases without SGLA or SSL (or neither).
However, it should be noted that in the 2D case we only consider
a slice of the 3D volume. Much of the
anatomical context is therefore missing, making it more difficult
to predict changes that are influenced by tissue
not shown on the slice in question.
For this purpose, we train our INR model on the 3D volumes as a whole
and perform the same ablation of SGLA and SSL again. 
The results are reported in \tableref{table:scores}. The difference between the different models in terms of PSNR, SSIM and MAE is relatively small in absolute terms,
because the image background is black, which is easy to predict
for a model. In other words, if the background remains constant and only the 
foreground improves, this reduces the effect on these scores.
However, we can still see that both SSL and SGLA improve all three scores.
Furthermore, even in the presence of a number of subjects with multiple scans available,
SGLA can be used in conjunction with SSL and still improve the performance,
in 2D as well as in 3D.

In addition to the three metrics we use in the 2D case,
in the 3D case, we can compute the head circumference (HC) based on
our predictions (details in \appendixref{appendix:circumference}).
Along with several other measurements, the HC is one of the
most important factors in determining prenatal development \cite{kiserud2017world}.
Therefore, we used it to measure how well our proposed method 
performs concerning the disentanglement of the subjects' age.
We compare the measured HC of our predictions
with the HC reported in the dHCP dataset, which we consider 
the ground truth. \tableref{table:scores}
displays the standard deviation $\sigma$ (in cm) of the error
between the measured HC of our prediction and the ground truth.
Furthermore, we report the correlation coefficient $r$ between
the predicted HC and the ground truth HC. 
Notably, we can see that with neither SGLA nor SSL,
the HCs of the predictions are almost uncorrelated with the ground truth HCs.
Again, our proposed INR with SGLA and SSL outperforms
the other ablated models in each of these measures or performs at least as well.

\begin{figure}[htbp]
\caption{Ablation of our model w.r.t. the HC. The blue circles show the HC ground truth,
  while the red x-es show the HC of our models' predictions on the test set.
}
\label{fig:ablationhc}
    \begin{subfigure}[b]{0.24\textwidth}
        \includegraphics[width=1\linewidth, clip, trim=4mm 4mm 3mm 3mm]{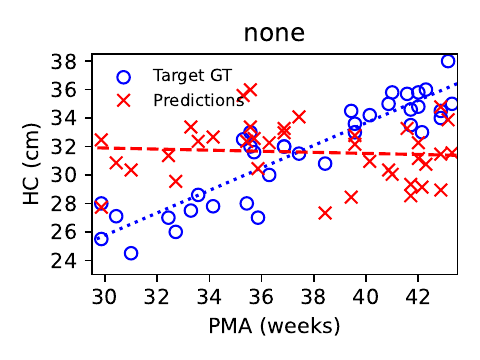}
    \end{subfigure}
    \begin{subfigure}[b]{0.24\textwidth}
        \includegraphics[width=1\linewidth, clip, trim=4mm 4mm 3mm 3mm]{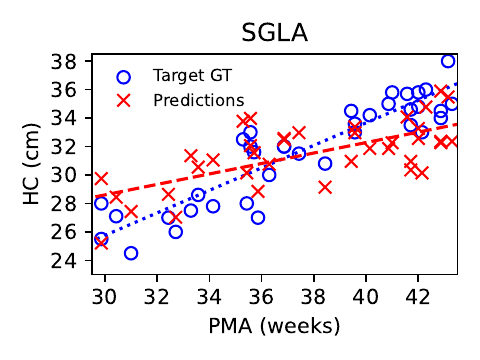}
    \end{subfigure}
    \begin{subfigure}[b]{0.24\textwidth}
        \includegraphics[width=1\linewidth, clip, trim=4mm 4mm 3mm 3mm]{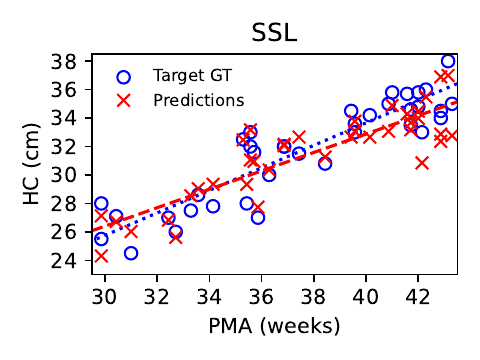}
    \end{subfigure}
    \begin{subfigure}[b]{0.24\textwidth}
        \includegraphics[width=1\linewidth, clip, trim=4mm 4mm 3mm 3mm]{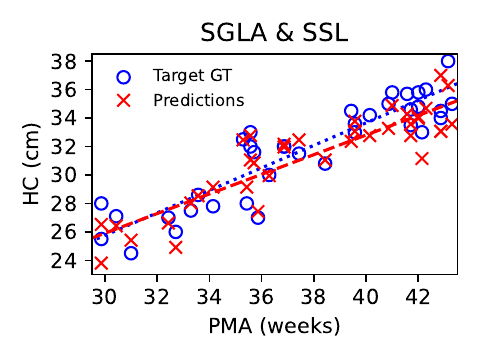}
    \end{subfigure}
      \vspace{-0.6cm}
\end{figure}

\begin{table}[htbp]
\caption{Average scores over the test set for the 2D INRs and the 2D DDM+GG baseline,
as well as for the 3D INRs.}
\label{table:scores}
\scalebox{0.8}{
\begin{tabular}{c|cc||ccc||ccc|cc}
 \multicolumn{3}{c||}{} & \multicolumn{3}{c||}{2D} & \multicolumn{5}{|c}{3D} \\
Model & SGLA & SSL & PSNR & SSIM & MAE & PSNR & SSIM & MAE & HC $\sigma$ & HC $r$\\ \hline
 \multirow{4}{*}{INR} & n & n & 19.5$\pm$2.6 & 0.642$\pm$0.058 & 0.0701$\pm$0.0178 &  22.9$\pm$3.1 & 0.804$\pm$0.036 & 0.0352$\pm$0.0091 & 3.850 & 0.173 \\
  & y & n & 20.6$\pm$2.5 & 0.683$\pm$0.065 & 0.0571$\pm$0.0146 &  23.8$\pm$2.8 & 0.832$\pm$0.049 & 0.0274$\pm$0.0083 & 2.137 & 0.835 \\
  & n & y & 21.3$\pm$2.1 & 0.715$\pm$0.075 & 0.0489$\pm$0.0124 &  24.5$\pm$2.4 & 0.850$\pm$0.054 & 0.0232$\pm$0.0079 & 1.071 & 0.962 \\
& y & y & {\bftab 21.4}$\pm$2.0 & {\bftab 0.722}$\pm$0.075 & {\bftab 0.0468}$\pm$0.0129 &  {\bftab 24.6}$\pm$2.4 & {\bftab 0.853}$\pm$0.055 & {\bftab 0.0225}$\pm$0.0078 & \bftab 0.964 & \bftab 0.968 \\ \hline
DDM+GG               & - & - & 19.3$\pm$2.1 & 0.632$\pm$0.088 & 0.0682$\pm$0.0158 & - & - & - & - & - \\
\end{tabular}
}
\end{table}
To better understand these results, we plot 
the ground truth HC and the measured HC of our predictions in 
\figureref{fig:ablationhc} for all four INRs.
We can see that the INR with neither SGLA nor SSL
completely fails to capture the developmental process.
Using either SGLA or SSL significantly improves the correlation
between the HC of the prediction and the ground truth HC. 
Finally using both jointly enables the model to capture the development process even better.

\begin{figure}[htp]
\begin{center}
\scalebox{0.5}{
\begin{tabular}{c@{}c@{}c@{}c@{}c@{}c@{}c@{}c@{}c@{}c@{}c} 
$t_1$, $t_2$ & \rotatebox{0}{\footnotesize Input} & \rotatebox{0}{\footnotesize Reconstruction} & \rotatebox{0}{\footnotesize Target GT} & \rotatebox{0}{\footnotesize Prediction} & \hphantom{xxxxxx<xx} & $t_1$, $t_2$ & \rotatebox{0}{\footnotesize Input} & \rotatebox{0}{\footnotesize Reconstruction} & \rotatebox{0}{\footnotesize Target GT} & \rotatebox{0}{\footnotesize Prediction} \\ 
$\begin{array}{c}29.9\\38.4\end{array}$ & \raisebox{-0.5\totalheight}{\includegraphics[width=0.17\linewidth]{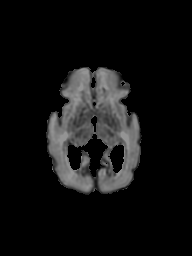}} & \raisebox{-0.5\totalheight}{\includegraphics[width=0.17\linewidth]{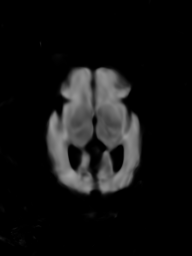}} & \raisebox{-0.5\totalheight}{\includegraphics[width=0.17\linewidth]{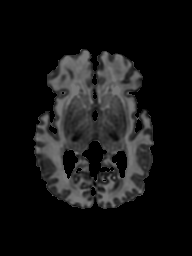}} & \raisebox{-0.5\totalheight}{\includegraphics[width=0.17\linewidth]{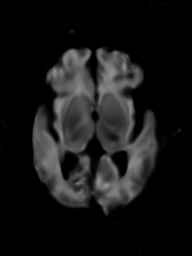}} & & $\begin{array}{c}31.0\\41.7\end{array}$ & \raisebox{-0.5\totalheight}{\includegraphics[width=0.17\linewidth]{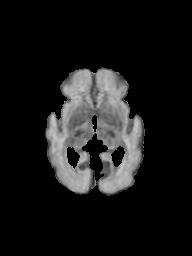}} & \raisebox{-0.5\totalheight}{\includegraphics[width=0.17\linewidth]{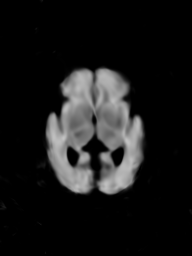}} & \raisebox{-0.5\totalheight}{\includegraphics[width=0.17\linewidth]{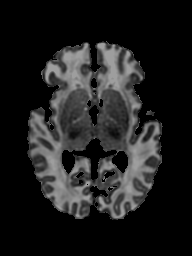}} & \raisebox{-0.5\totalheight}{\includegraphics[width=0.17\linewidth]{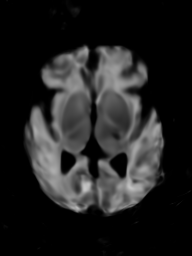}} \\ 
$\begin{array}{c}36.3\\42.9\end{array}$ & \raisebox{-0.5\totalheight}{\includegraphics[width=0.17\linewidth]{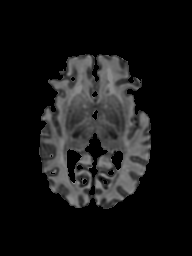}} & \raisebox{-0.5\totalheight}{\includegraphics[width=0.17\linewidth]{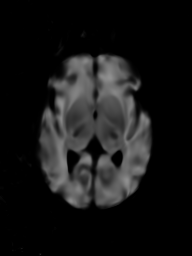}} & \raisebox{-0.5\totalheight}{\includegraphics[width=0.17\linewidth]{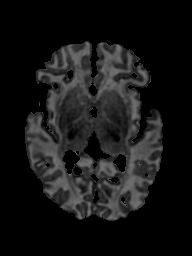}} & \raisebox{-0.5\totalheight}{\includegraphics[width=0.17\linewidth]{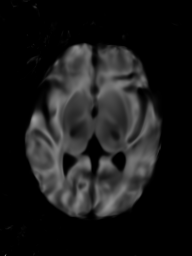}} & & $\begin{array}{c}43.1\\35.3\end{array}$ & \raisebox{-0.5\totalheight}{\includegraphics[width=0.17\linewidth]{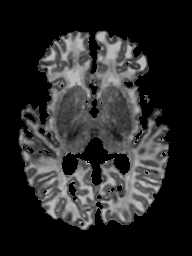}} & \raisebox{-0.5\totalheight}{\includegraphics[width=0.17\linewidth]{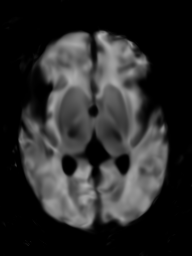}} & \raisebox{-0.5\totalheight}{\includegraphics[width=0.17\linewidth]{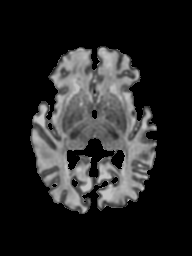}} & \raisebox{-0.5\totalheight}{\includegraphics[width=0.17\linewidth]{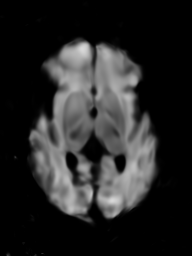}} \\ 
$\begin{array}{c}33.6\\40.9\end{array}$ & \raisebox{-0.5\totalheight}{\includegraphics[width=0.17\linewidth]{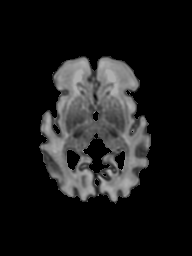}} & \raisebox{-0.5\totalheight}{\includegraphics[width=0.17\linewidth]{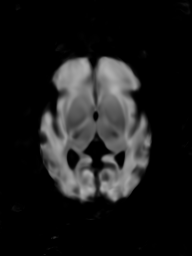}} & \raisebox{-0.5\totalheight}{\includegraphics[width=0.17\linewidth]{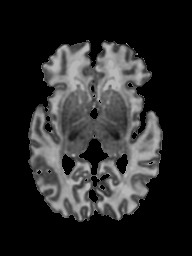}} & \raisebox{-0.5\totalheight}{\includegraphics[width=0.17\linewidth]{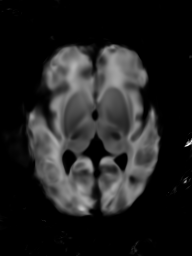}} & & $\begin{array}{c}42.9\\35.6\end{array}$ & \raisebox{-0.5\totalheight}{\includegraphics[width=0.17\linewidth]{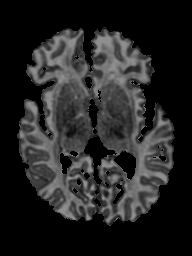}} & \raisebox{-0.5\totalheight}{\includegraphics[width=0.17\linewidth]{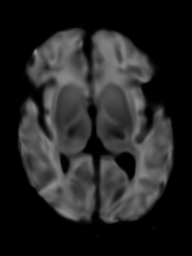}} & \raisebox{-0.5\totalheight}{\includegraphics[width=0.17\linewidth]{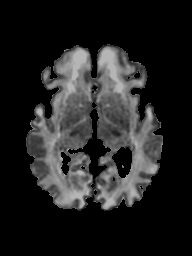}} & \raisebox{-0.5\totalheight}{\includegraphics[width=0.17\linewidth]{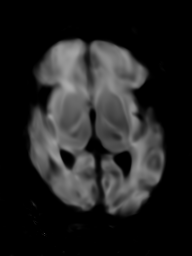}}
\end{tabular}
}%
\end{center}
\caption{Six examples from the test set, along with the PMA $t_1$ of the input, and the PMA $t_2$ of the target ground truth image.}
\label{fig:quadruples}
\end{figure}

\paragraph{Qualitative Results}
In \figureref{fig:quadruples}, we show two examples from the test set.
We show four images per subject: (1) the \emph{input} \scalebox{0.9}{\smash{$\vec I^1$}}
with its age $t_1$
that gets encoded into a latent vector, (2) the 
reconstruction \scalebox{0.9}{\smash{$\hat{\vec I}^1$}} of the input at $t_1$,
(3) the
\emph{target ground truth} \scalebox{0.9}{\smash{$\vec I^2$}} from $t_2$ and (4) the prediction \scalebox{0.9}{\smash{$\hat{\vec I}^2$}} at age $t_2$.
Note that for our model, $t_2$ does not necessarily have to be greater
than $t_1$, we can also choose to look back in time to predict
what a given brain has looked like in the past.
The reconstructions display the input images with 
some loss of details due to the bottleneck created
by the low-dimensional latent space.
The predictions generally match the size and 
the contrast of the target ground truth
well, but have trouble predicting the exact shape of the cortical folds.
This is, however, a difficult task, as even for monozygotic twins, 
the folds exhibit individual patterns \cite{white2010development}.
On the bottom right, we see an interesting example where the
mid-sagittal plane of the input is slightly off the vertical,
in contrast to the target ground truth. Interestingly, however, 
the reconstruction, as well as the prediction, display this slight
rotation, which means it must have been encoded in the latent
vector.

\begin{figure}[htbp]
\begin{center}
  \scalebox{1.00}{
    \begin{subfigure}[b]{0.5\textwidth}
        \includegraphics[width=0.9\linewidth, clip, trim=4mm 3mm 3mm 3mm]{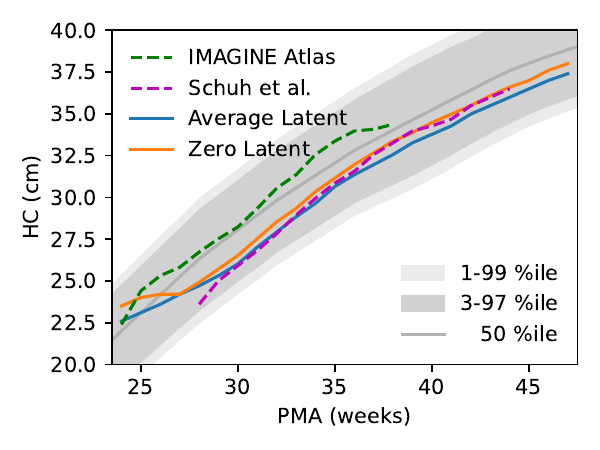}
        \phantomsubcaption
        \label{fig:imagine:hc}
    \end{subfigure}
    \begin{subfigure}[b]{0.5\textwidth}
        \scalebox{0.65}{
            \begin{tikzpicture}
            \node at (0,0) [anchor=south west]{\includegraphics[width=5cm, clip, trim=0 0 0 0]{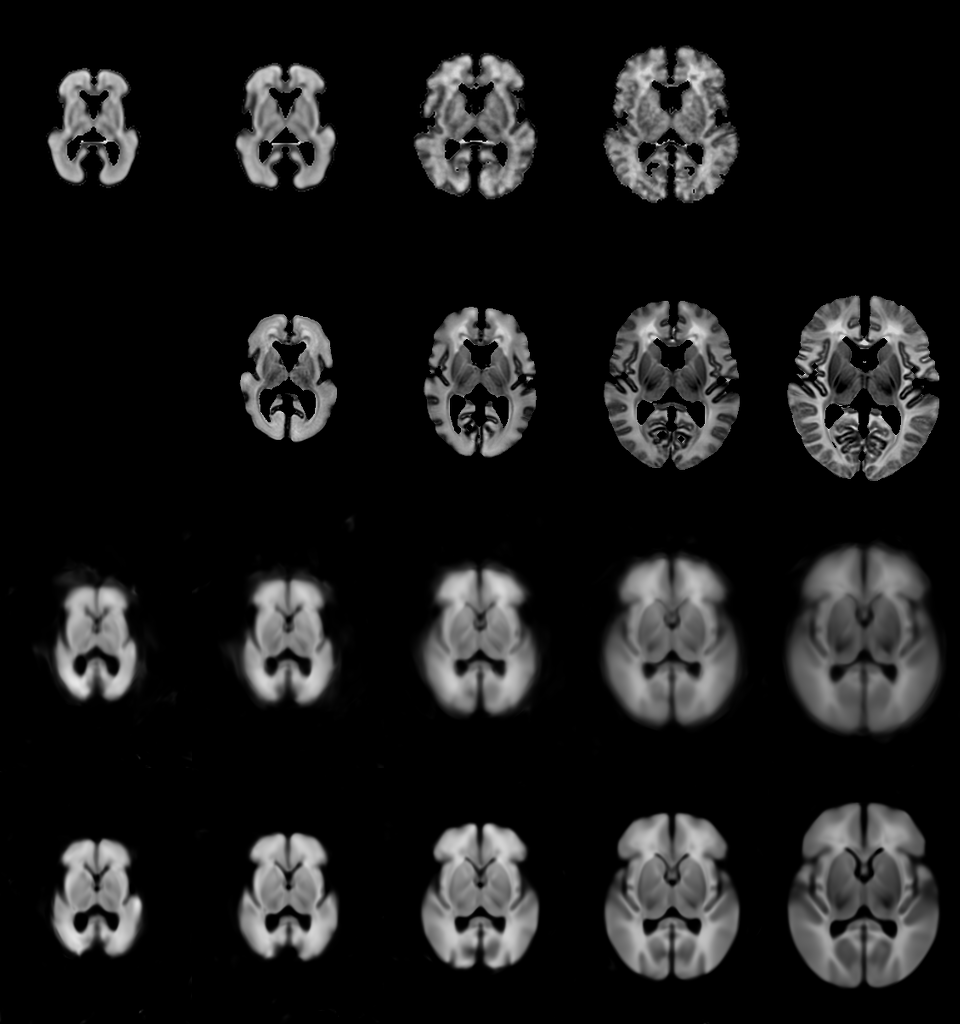}}; %
            \foreach \i/\age in {0/26, 1/29, 2/33, 3/38, 4/44}{
                \node at (0.6+\i, 0.2)[anchor=north, scale=0.8] {\vphantom{$2^2$}\age};
            }
            \node at (0.2, 4.2) [anchor=south west, rotate=90, align=left, scale=0.75] {\vphantom{Ig}IMAGINE      \\ \vphantom{Ig}Atlas};
            \node at (0.2, 2.8) [anchor=south west, rotate=90, align=left, scale=0.75] {\vphantom{Ig}Schuh et al. \\ \vphantom{Ig}Atlas};
            \node at (0.2, 1.4) [anchor=south west, rotate=90, align=left, scale=0.75] {\vphantom{Ig}Zero         \\ \vphantom{Ig}Latent};
            \node at (0.2, 0)   [anchor=south west, rotate=90, align=left, scale=0.75] {\vphantom{Ig}Average      \\ \vphantom{Ig}Latent};
            \node at (0, -0.45) {$ $};
            \end{tikzpicture}
        }
        \phantomsubcaption
        \label{fig:imagine:imgs}
    \end{subfigure}
    }
    \end{center}
\caption{Comparison of our ``average'' brain development
  with the IMAGINE atlas and the atlas of Schuh et al. in terms of HC growth curves \figureref{fig:imagine:hc}, along with the
  quantiles reported in \cite{fenton2003new}.
  Furthermore, we show the corresponding axial slices in \figureref{fig:imagine:imgs}.
}
\label{fig:imagineatlas}
\end{figure}
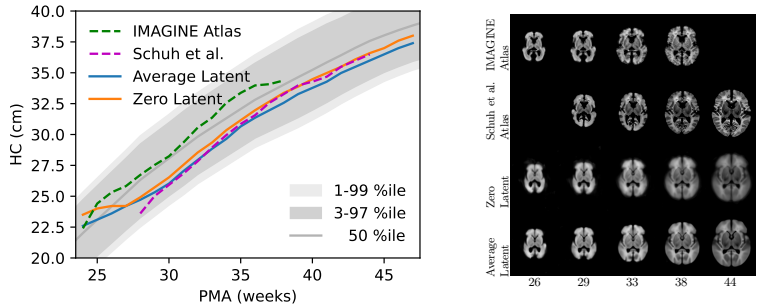

\paragraph{Predicted Average Development}
To gain insight into the development process 
that the trained network captured, 
we consider two methods of extracting an
``average'' brain for each age, similar to creating an age-resolved atlas.
Firstly, since we have designed the latent space
to represent the characteristics of a subject, it can be expected that
averaging the latent vectors of a population (i.e., our training set)
would result in an average-looking brain. 
Secondly, because we initialize the latent vectors with a zero-vector, it is also
possible to use it for the generation of an ``average'' brain.
Thus, we use these two (\emph{zero-} and \emph{average-}) latent vectors 
to generate a temporal sequence of images of the trained model with SSL and SGLA.
Furthermore, we compare the two approaches with the
IMAGINE fetal atlas \cite{gholipour2017normative}, %
as well as the neonatal atlas of Schuh et al.\ \cite{schuh2018unbiased}. 
The latter is also based on the dHCP dataset.
In \figureref{fig:imagineatlas}, we report the measured HCs 
along with percentiles of a
fetal-infant preterm growth chart \cite{fenton2003new,fenton2007using}.
For instance, for diagnosing microcephaly, bounds like 
the 1\textsuperscript{st} or 3\textsuperscript{rd} percentile 
or three standard deviations of the HC are being used \cite{desilva2017congenital}.
Under these criteria, all four growth curves are within the normal range.
Notably, the HC lines of our generated ``average''-brains agree very well
with the atlas of \cite{schuh2018unbiased}, which is based on the same dataset.
The IMAGINE atlas \cite{gholipour2017normative},
exhibits slightly larger HCs, but is created using a set of \emph{fetal}
MR-images, with a significantly smaller sample size of 81 scans.
However, it also remains within the normal interval.
The quality of the two generated sequences using the zero- and average latent vectoris different:
The series generated by the average latent vector is more detailed and has less blurry features.

\paragraph{Computational Resources}\label{sec:resources}
In \tableref{table:resources}, we report the time and GPU memory 
used. We used the same settings regarding 
pixel-sampling and micro-batching 
for the 2D- and 3D case.
As elaborated in \sectionref{sec:pixelsampling}, 
we can reduce the number of pixels sampled
per optimization step. Alternatively, we have the option
to use micro-batching to trade memory for time.
Finally, the number of parameters used in the 2D and 3D INRs
is more than two orders of magnitude smaller than in
the 2D DDM+GG baseline.

\begin{table}
\caption{
Resources used for the INRs in 2D and 3D and the DDM+GG baseline.
}
\label{table:resources}

\begin{center}
\scalebox{0.8}{
\begin{tabular}{l|rr|rr|r}
& \multicolumn{2}{c}{GPU Memory (in MB)} & \multicolumn{2}{c}{Time} \\ \
Model & Training & Inference & Training (total) & Inference (per Sample) & \# Parameters \\ \hline
INR 3D  & 10350 & 10286 &  81.6\,h & 5.20\,min & $232\;\;450$ \\ \hline
INR 2D  &   284 &   232 &   2.5\,h & 0.73\,min & $232\;\;194$ \\ \hline
DDM 2D &  7412 &  \multirow{2}{*}{1332} &  76.8\,h & \multirow{2}{*}{1.80\,min} & $113\;\;669\;\;762$   \\
Regression 2D  &  6412 &      &   7.9\,h &  & $86\;\;786\;\;817$ \\
\end{tabular}
}
\end{center}
\end{table}

\section{Conclusion}
We present a method for modelling the neonatal brain development using INRs.
We show that it is necessary to disentangle the subject's identity from
its age. We propose and evaluate two novel methods, SSL and SGLA.
We implement them in 2D as well as in 3D. On the one hand, we compare the predictions with the ground
truth with respect to image quality, and on the other hand also indirectly via
the HC as an important development metric. We demonstrate how our two proposed solutions improve the results through the disentanglement.
However, we note that the image synthesis with INRs still has open challenges  
such as the image quality, and the modelling of processes like the cortical folding, which
are additionally influenced by other factors than time. We anticipate that
larger datasets would enable more expressive networks to be trained,
which could in turn alleviate these issues to some degree.
To avoid the need for coregistration of the scans as a preprocessing step,
it would be interesting to extend the disentanglement also
to geometric transformations.
In future work, we would like to explore extending the model to predict
segmentations, which could be used to generate actual atlases.

\begin{credits}
\subsubsection{\ackname}

We are grateful for the support of 
the Novartis FreeNovation
initiative and the Uniscientia Foundation (project \#147-2018).
We would also like to thank the NVIDIA Corporation for donating a GPU that was used for our experiments.

Data were provided by the developing Human Connectome Project, KCL-Imperial-Oxford Consortium funded by the European Research Council under the European Union Seventh Framework Programme (FP/2007-2013) / ERC Grant Agreement no. [319456]. We are grateful to the families who generously supported this trial.

\subsubsection{\discintname}
The authors have no competing interests to declare that are relevant to the content of this article.
\end{credits}
\bibliographystyle{splncs04}
\bibliography{prime}

\newpage
\appendix 

\section{Detailed Architecture}\label{appendix:detailedarchitecture}
Our network $f_\theta$ is built using affine blocks with
WIRE \cite{saragadam2023wire} activations which are state-of-the-art for INRs,
and have been shown to outperform the
previously state-of-the-art SIREN networks \cite{sitzmann2020implicit} 
in many applications of INRs. They have the advantage of avoiding
the additional complexity of positional encodings.
We performed experiments using different residual topologies. 
Based on the subjective quality of the generated images on the training set, 
we chose to use a fully residual architecture, 
that also has been proposed in \cite{stolt2023nisf}.
Specifically, we use a residual architecture with additive skip connections around each 
WIRE block with the same input and output dimension.
Furthermore, we chose a network depth of nine layers,
based on an ablation over the number of layers
and the subjective appearance of the image quality during training.

More specifically, our network $f_\theta$ is composed of multiple layers $\phi_i$.
Let $h \in \mathbb N$ be the number of hidden dimensions, $\lambda \in \mathbb N$ the latent dimension
and $d$ the number of spatial dimensions. The INR network is defined as
\begin{equation}
    f_\theta = \phi_N \circ \psi_{n-1} \circ \ldots \circ \psi_2 \circ \phi_1
\end{equation}
with residual blocks $\psi_i(x) = \frac12(x + \phi_i(x))$, similar as in \cite{resnet2016}. Furthermore, the WIRE
blocks are
\begin{equation}
    \begin{array}{rrl}
    \phi_i : & \mathbb R^{n_i} \to & \mathbb R^{m_i} \\
             & x \mapsto & \sin(\omega_0 u_i(x)) e^{-(s_0 v_i(x))^2} 
    \end{array}
\end{equation}
where $u_i, v_i: \mathbb R^{n_i} \to \mathbb R^{m_i}$ are affine maps with trainable parameters.
The application of the scalar functions to vectors is understood to be element-wise.
In our case $\lambda = m_1 = \ldots m_{N-1} = n_2 = \ldots = n_N = h = 128$, $m_N = 1$ and $n_1 = \lambda +d+1$
and $d=2$ for the 2D case, and $d=3$ for the 3D case.

\section{Hyperparameter Optimization DDM+GG}\label{appendix:ddim}
\begin{figure}[htbp]
\begin{center}
\includegraphics[width=0.6\linewidth, clip, trim=0mm 0mm 0mm 0mm]{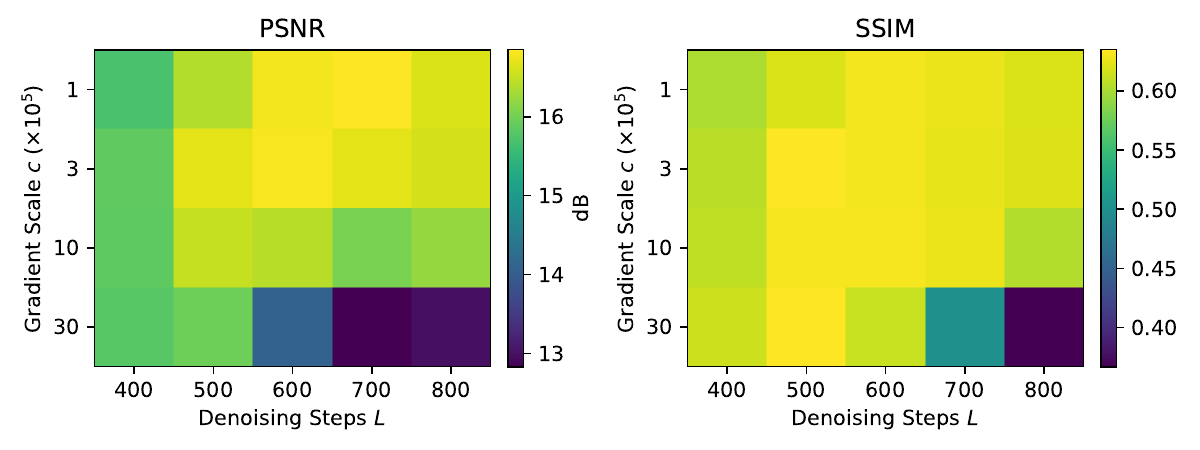}
\caption{PSNR and SSIM on the test set as a function of the
  number of noising and denoising steps $L$ and the gradient scale $c$ for the DDM+GG baseline \cite{wolleb2022swiss}.}
\label{fig:diffusion_parameters_quality}
\end{center}
\end{figure}

To allow for a fair comparison, we optimized the parameters
of the DDM+GG sampling on the test set via a grid search.
First, we performed a visual inspection over a coarse grid of
samples with $c \in [10^5, 10^8]$ and $L \in [100, 700]$.
Then, we performed a finer grid search and evaluated 
the metrics described in \sectionref{sec:experiments}.
The metrics are shown in \figureref{fig:diffusion_parameters_quality}. 
We chose $L=600$ and $c=3\cdot 10^5$, as the PSNR metric
was maximal in this case. The maximal SSIM was attained at $L=500$ with the same value of $c$,
but overall, the SSIM metric seems to be a lot less sensitive 
to these two hyperparameters, which is why we prioritized 
the PSNR metric for this choice.

\section{Preprocessing} \label{appendix:preprocessing}
We aligned the central (mid-sagittal) plane to the image axes. 
The mid-sagittal plane was determined by rigidly registering 
the images to their mirror image, as proposed in \cite{tuzikov2002brain}.
Then, we adjusted the orientation by registering the image
to an average over the dataset with only intra-plane translations,
rotations, and isotropic scaling. Finally, we undo the
scaling from the registration to get back to the original size.

We performed a skull-stripping that also removed
all the cerebrospinal fluid (CSF), 
to avoid unnecessary hyperintense regions.
The background voxels were set to zero, and the range 
between the first and 99th percentile of the foreground was normalized
to $[0,1]$.
For the training, we used 329 subjects with only a single scan
and 19 subjects with two scans. For the validation, 
we used four subjects with two scans; for testing, we used 
another 19 subjects with two or more scans.
After the preprocessing, we are left with volumes of dimension 
$192 \times 256 \times 192$ with an 
isotropic voxel size of $0.588\text{ mm}$.
We normalize the spatial input coordinates to $[-0.5, 0.5]$,
and divide the temporal coordinates (i.e.\ PMA in weeks) by $100$
to map them into the range $[0,1]$.

\section{Hyperparameters of the INRs}
\subsection{Ablation of $p$ parameter in SGLA}\label{appendix:ablationp}
\begin{figure}[htbp]
\begin{center}
\includegraphics[width=0.5\linewidth, clip, trim=0mm 0mm 0mm 0mm]{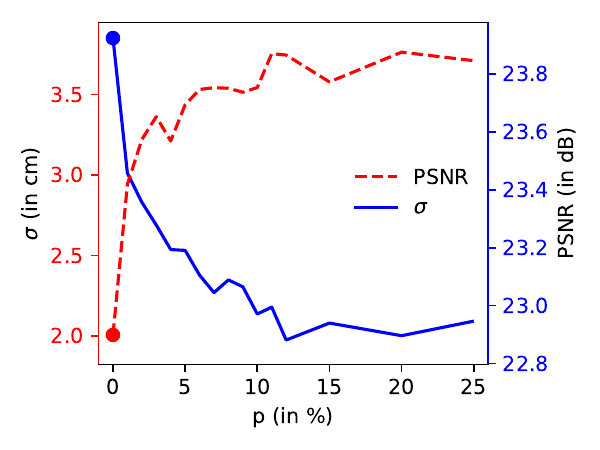}
\end{center}
\caption{Ablation of the probability parameter $p$ for SGLA. We report $\sigma$ and the PSNR as in 
\sectionref{sec:experiments}. The dots indicate $p = 0$ i.e. when SGLA is not used. 
}
\label{fig:ablationp}
\end{figure}

As outlined in \sectionref{sec:method} we chose $p=10\%$. To verify if this is optimal,
we perform an ablation with respect to $p$ in the range $0\%$ to $25\%$ and 
report $\sigma$ and PSNR as in \sectionref{sec:experiments} in \figureref{fig:ablationp}.
As already reported in \sectionref{sec:experiments}, SGLA with $p=10\%$ improves the predictions
over the baseline without SGLA. In \figureref{fig:ablationp} we see that it would probably be possible to 
marginally increase the performance, but for $p > 10\%$ the improvements start to diminish. 

\subsection{Foreground- to Background Ratio}\label{appendix:pixelsampling}
If we were to sample the pixels uniformly from the given image during training, 
we would get a proportionally large amount of background pixels. 
We found that the uniform background is easy to learn for the network,
while the foreground is more challenging.
We therefore introduced a second hyperparameter, that sets a fixed ratio
of foreground pixels to background pixels sampled per step. 
We set this ratio to $90:10$ to favor the foreground.%
In the inference we use the same foreground to background ratio as in the training.
For the 2D case we could afford to use all pixels in each optimization step of the inference.

\section{Linking Head- \& Brain Circumference}\label{appendix:circumference}
As we performed skull-stripping as a preprocessing step, 
it is not possible to measure the head circumference (HC) directly.
The HC is usually estimated by measuring it with a flexible tape measure, 
or by measuring the occipitofrontal and biparietal diameters to define an ellipse. 
The HC is then estimated using the circumference \cite{salomon2019isuog} of said ellipse.
Alternatively, it can also be measured using a flexible tape measure. In both cases, not choosing the correct plane in
which the HC is measured can lead to errors. %

\begin{figure}[htbp]
\begin{center}
\includegraphics[width=0.5\linewidth, clip, trim=0mm 0mm 0mm 0mm]{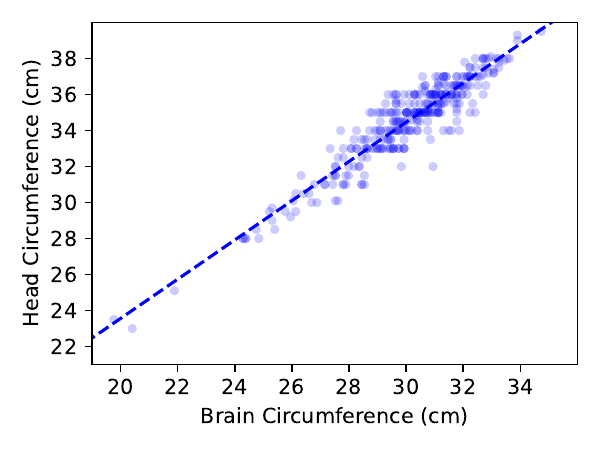}
\end{center}
\caption{Linear Regression between Brain- and Head circumference: 
  $HC=1.090BC+1.758\text{ cm}$, $\sigma=0.8600\text{ cm}$, $r=0.9358$}
\label{fig:head_circumference_prediction}
\end{figure}

We employ the latter method to get the brain circumference (BC)
and perform a linear regression between measured BC
and reported HC on the training set.
As shown in \figureref{fig:head_circumference_prediction}, these
two values are highly correlated ($r=0.9358$). %
The error has a standard deviation of $\sigma = 0.86\text{ cm}$.
We use this linear model to predict the HC in our experiments, which
means that this value should be considered a lower bound.

With this indirect method, a source of error is the
relationship between BC and HC. Due
to variations in thickness of the skull and the surrounding
tissue, the linear model does not capture this relationship perfectly. 
Furthermore, as can be seen from the vertical clustering in 
\figureref{fig:head_circumference_prediction}, some measurements appear more frequently than others.
We estimate that about half of the measurements was rounded to $5 \text{ mm}$ and about one third to  $10 \text{ mm}$,
while the rest was rounded to $1 \text{ mm}$.

To justify this, we compare the accuracy of this linear model with
other errors reported in the literature, and point out some other error sources:
The intra-observer standard deviation for the ellipse based HC measurement is 
$0.615\text{ cm}$ \cite{sarris2012intra}, which is slightly lower but comparable to the
error we estimated for our method. It is, however, reported for sonography, and not for MRI,
and with manual placement of the ellipse, but it still allows us to get a rough
estimate of the error we can expect.

\end{document}